# Properties of the skyrmion crystal in the Heisenberg triangular lattice with scalar chirality

H. Bocquet [1,*] C. J. Ganahl [2,†] M. S. Scheurer [3,‡] P. M. Derlet,[1,§] and A. M. Läuchli[1,4]
[1] PSI Center for Scientific Computing, Theory and Data, 5232 Villigen PSI, Switzerland
[2] Institut für Theoretische Physik, Universität Innsbruck, Technikerstr. 21A, A-6020 Innsbruck, Austria
[3] Institut für Theoretische Physik III, Universität Stuttgart, Pfaffenwaldring 57, 70550 Stuttgart, Germany
[4] Institute of Physics, Ecole Polytechnique Fédérale de Lausanne (EPFL), 1015 Lausanne, Switzerland

Skyrmion crystals have been primarily discovered under a magnetic field for materials with noncentrosymmetric interactions. More recent developments have investigated the stability of skyrmion crystals in itinerant magnets without a magnetic field. In this study, we find that a type of skyrmion crystal with two topological charges per unit cell and no magnetization at the ferromagnetic point in reciprocal space, SkX-2, is naturally stabilized in an SO(3)-symmetric model with short-range interactions realized by the Heisenberg model on the triangular lattice with scalar chirality. We complement our numerical results with a theoretical analysis that quantitatively describes the transition from the ferromagnetic to the SkX-2 ground state and the evolution of the topological charge density. Despite the constraints given by the Mermin-Wagner theorem at finite temperature, the SkX-2 can either exhibit a first-order phase transition with translation symmetry breaking, or a continuous transition to a floating solid depending on the charge density, which is given by the strength of the scalar chirality interaction. Finally, the numerical results indicate the presence of $\mathbb{Z}_2$-vortices at finite temperature above the tetrahedral ground state in the antiferromagnetic region of our model, suggesting the existence of a vortex topological transition similar to that reported in the ordering of the pure antiferromagnetic model.

## I. INTRODUCTION

Since the experimental discovery of the first magnetic skyrmion crystal in MnSi [1], there is growing interest in understanding and classifying skyrmion textures, in order to identify their existing conditions and their full possible technological advantages [2,3]. The simplest skyrmion lattice, accounting for the description of several materials [1,4–6], is composed of distinct skyrmions defined by a topological charge $Q = \pm 1$ and stabilized by the presence of a noncentrosymmetric Dzyaloshinskii-Moriya interaction (DMI) and a magnetic field. Skyrmions with $Q = 1$ or antiskyrmions with $Q = -1$ are preferred depending on the sign of the magnetic field and the DMI. In contrast, skyrmion arrangements predicted in frustrated models without DMI can have either a positive or a negative charge [7–9]. This is due to the remaining spin reflection symmetry in the plane perpendicular to the field, which can only be spontaneously broken. An interesting and experimentally relevant extension consists of the addition of easy-plane anisotropy, which effectively increases the packing fraction of skyrmions to the point where fractional amounts of charge can localize [10]. The result could be a texture composed of merons with $Q = \pm 1/2$, which, unlike skyrmions that embody a full magnetization flip between the core and the perimeter, exhibit in-plane magnetization at the perimeter [11–13].

Another important situation to consider is when no magnetic field is applied. Topological magnetic textures without global magnetization have been theoretically proposed in this context for itinerant magnets, where the effect of conduction electrons on local magnetization can be described using the Kondo-lattice model [14–16], or directly via a momentum-resolved exchange term mediating the effective long-range interaction between local magnetization, such as the Ruderman-Kittel-Kasuya-Yosida interaction [17–20]. These fundamental examples lay the groundwork for interpreting the field-free emergence of skyrmion crystals, for instance, in metallic $Gd_2PdSi_3$ [21], $Gd_3Ru_4Al_{12}$ [22], or $GdRu_2Si_2$ [23]. The existence of skyrmion crystals without an external magnetic field is also expected in the presence of a compass interaction [19,24], or in a multilayer model [25]. Among these many different examples, skyrmions arrange to form various crystals exhibiting $Q = \pm 1$ skyrmions, but also merons, and more complex distributions of the topological charge. Thus one method to simply classify these textures is to count the topological charge per unit cell.

In the present study, we are interested in a model that has both short-range interactions and retains the three-dimensional rotation symmetry of the exchange interaction. We consider a classical Heisenberg Hamiltonian augmented with a scalar chirality interaction on the triangular lattice. The latter can arise due to the effect of the magnetic field on the electron hopping in the corresponding tight-binding model [26] and, in the quantum regime, supports the emergence of a chiral spin liquid, a phase characterized by topological order, long-range entanglement, and broken time-reversal symmetry [27–30]. In the classical regime with a ferromagnetic exchange, we find that this model admits a skyrmion crystal, also called SkX-2 in the following, composed of two charges per unit cell in the ground state and observed in more complex itinerant models in Refs. [14,17]. Unlike its more common counterpart with a unique topological charge per unit cell (SkX-1), the charge density distribution in SkX-2 is not localized at unit-charge centers that can be directly associated with individual skyrmions. Under an applied magnetic field, however, this charge can continuously redistribute until a discontinuous phase transition to SkX-1 occurs, which eventually relocalizes the charge into conventional skyrmion cores [14]. In our model, we will predict analytically the exact evolution of the SkX-2 topological charge density, often an elusive question relegated to numerical computation. This will simultaneously justify the existence of a continuous transition between the ferromagnetic state (stabilized at large ferromagnetic exchange) and the SkX-2 state.

*Contact author: h.bocquet@protonmail.ch
†Contact author: clemens.ganahl@student.uibk.ac.at
‡Contact author: mathias.scheurer@itp3.uni-stuttgart.de
§Contact author: peter.derlet@psi.ch

In spin space, the scalar chirality, similar to the skyrmion crystal supported by DMI instead of frustration in the presence of a magnetic field, breaks the $\mathbb{Z}_2$ symmetry related to the sign of the topological charge. At finite temperature, the remaining SO(3) symmetry provides no support for spontaneous symmetry breaking for this two-dimensional model [31,32]. Nevertheless, depending on the charge density, we will demonstrate that SkX-2 can exhibit either a first-order phase transition or a liquid-to-floating-solid transition as predicted in Kosterlitz-Thouless-Halperin-Nelson-Young (KTHNY) theory for two-dimensional particle systems [32–34].

When the exchange interaction is antiferromagnetic, a tetrahedral order is stabilized over a finite range of the parameter space in addition to the usual 120°-order. Although both these phases cannot retain their order under thermal fluctuations, we will discover the presence of $\mathbb{Z}_2$ vortices on the tetrahedral order, suggesting the existence of a topological transition as already proposed for the 120°-order in Ref. [35].

The paper is organized as follows. In Sec. II, we introduce our augmented Heisenberg model mathematically and unveil the ground states numerically thanks to a modified basin-hopping procedure. The transition to the SkX-2 ground state is predicted analytically through a leading-order continuum expansion on the ferromagnet in Sec. III. At higher order in the expansion, we derive a Ginzburg-Landau model to describe the topological charge density evolution in Sec. IV. In Sec. V, we reveal the thermal properties of the SkX-2 using Monte Carlo simulations, demonstrating the existence of a floating solid and a long-range order at low temperature, respectively, at small and large charge density. We show in addition the presence of $\mathbb{Z}_2$ vortices at finite temperature on the tetrahedral order. Finally, we discuss the results in Sec. VI.

## II. MODEL AND GROUND STATES

We consider a classical Heisenberg model with a chiral interaction term, given by the Hamiltonian:

$$\mathcal{H} = J\sum_{\langle i,j\rangle} \boldsymbol{S}_i\cdot\boldsymbol{S}_j + J_\chi \sum_{(i,j,k)\in\Delta} \boldsymbol{S}_i\cdot(\boldsymbol{S}_j\times\boldsymbol{S}_k). \tag{1}$$

The spins $\boldsymbol{S}_i$ are unit vectors in the three-dimensional Euclidean space. For the triangular lattice, the chiral interaction is evaluated over all elementary triangles according to a clockwise permutation of the vertices $i$, $j$, and $k$. The favored chirality is determined by the sign of $J_\chi$. Although the Heisenberg interaction is by itself $O(3)$-symmetric, the present Hamiltonian is SO(3)-symmetric due to the chiral interaction. The spin reflection symmetry of the Heisenberg interaction corresponds to a sign change for the chiral interaction, i.e., a switch in the favored chirality. Consequently, changing the favored chirality leaves the physics invariant up to a reflection. With this in mind, we need only consider the Hamiltonian for $J_\chi > 0$. To simplify the description of the parameter space, we define the angle $\theta \in [0, \pi]$, such that $J = \cos(\theta)$ and $J_\chi = \sin(\theta)$.

To find the ground state, we use a modified basin-hopping [36] procedure, where we borrow ideas from genetic algorithms [37] to improve convergence. The starting point is a family of randomly generated candidate states. For each candidate, a Powell search [38] gives us local minima of the energy function. We then sort the candidates according to their energy and replace the worst candidates with new random states. The other candidates are modified by introducing random displacements (mutation) and crossovers of low-energy candidates, chosen randomly with a roulette wheel selection [39]. Furthermore, we decrease the magnitude of the displacements at each iteration step using a temperature parameter. Figure 1 shows the resulting ground-state energy as a function of $\theta$ and Fig. 2 illustrates the ground-state configurations. For small $\theta$, the ground state is the well-known 120°-AFM configuration [Fig. 2(a)], followed by the tetrahedral order [Fig. 2(b)] for $0.3140 \leqslant \theta \leqslant \pi/2$. In the large exchange limit, we find the expected ferromagnetic order for $\theta \geqslant 2.4190$.

In addition to these well-known orders, we identify a region where a skyrmion crystal SkX-2 is the ground state between the tetrahedral order at $\theta = \pi/2$ and the ferromagnetic order at $\theta = 2.4190$ by inspecting the structure factor. The SkX-2 is a multi-Q state, but unlike the usual SkX-1, it does not bear a magnetization at the ferromagnetic point in reciprocal space ($\Gamma$ point), which explains its relevance in the absence of a magnetic field. It is described up to a global rotation by the following modulation of the magnetization field:

$$\boldsymbol{m}(\boldsymbol{r}) \propto \begin{bmatrix} \cos(\boldsymbol{X}_2\cdot\boldsymbol{r}) + \cos(\boldsymbol{X}_3\cdot\boldsymbol{r}) \\ \cos(\boldsymbol{X}_2\cdot\boldsymbol{r}) - \cos(\boldsymbol{X}_3\cdot\boldsymbol{r}) \\ \frac{\sqrt{2}J_\chi}{|J_\chi|}\cos(\boldsymbol{X}_1\cdot\boldsymbol{r}) \end{bmatrix}. \tag{2}$$

The SkX-2 reciprocal vectors $\boldsymbol{X}_i$ are three complementary vectors of the same magnitude, such that $\boldsymbol{X}_1+\boldsymbol{X}_2+\boldsymbol{X}_3 =$

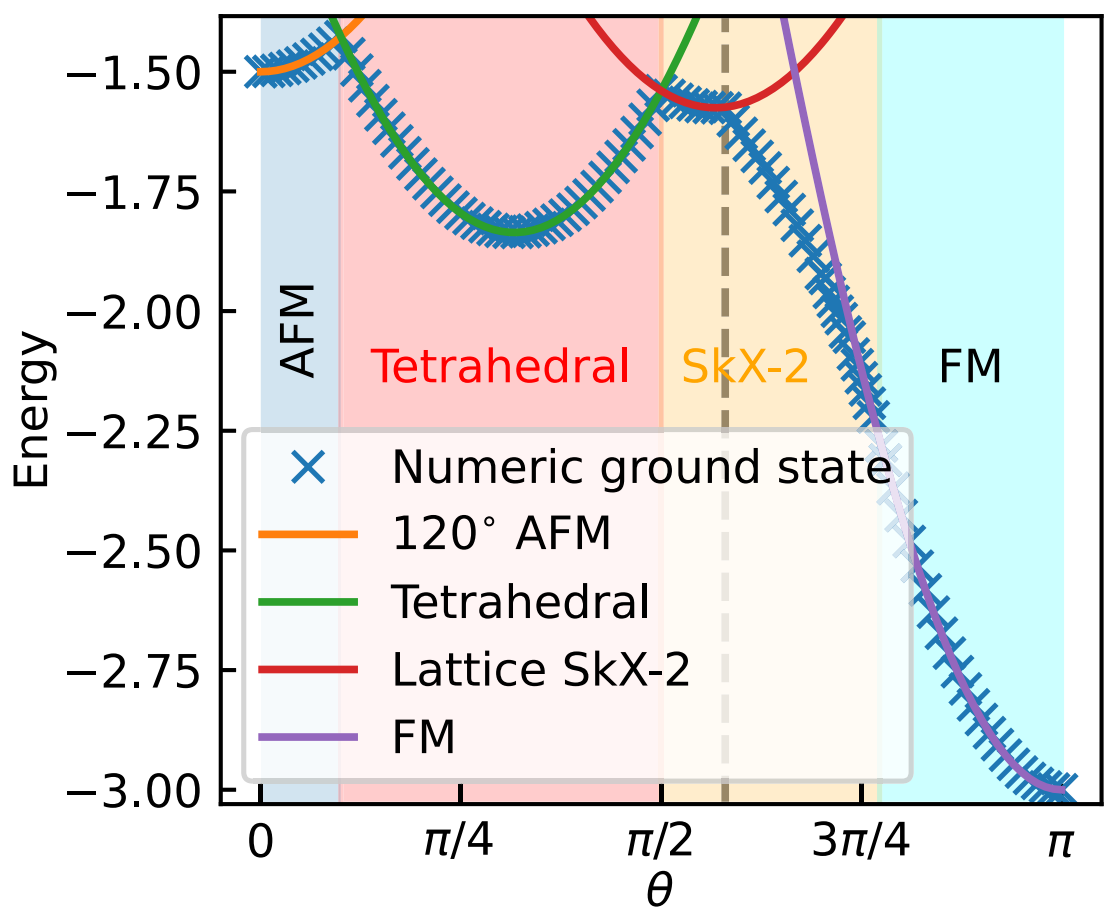

FIG. 1. Ground-state energy per spin of the Hamiltonian in Eq. (1) as a function of $\theta = \arctan(J_\chi/J)$. The solid lines represent analytical ground-state energy curves, and crosses are the numerical values obtained using a basin-hopping procedure for doubly periodic triangular spin lattices of various sizes (between $6 \times 6$ and $36 \times 36$ spins) to limit boundary effects (only minimal energy is provided). The background colors mark different ordered states, where the SkX-2 region contains both the lattice SkX-2 and the free SkX-2 separated by the vertical dashed line.

0. In a region close to the tetrahedral order, for $\pi/2 \leqslant \theta \leqslant 1.8221$, we find that these reciprocal vectors remain constant and commensurate with the triangular lattice reciprocal vectors: $\boldsymbol{X}_1 = \frac{2\pi}{3a}(1,0)$, $\boldsymbol{X}_2 = \frac{2\pi}{3a}(-1/2, \sqrt{3}/2)$ and $\boldsymbol{X}_3 = \frac{2\pi}{3a}(-1/2, -\sqrt{3}/2)$. This regime corresponds to the lattice limit, or the Nyquist limit, of the SkX-2, as the order on the underlying triangular lattice could not be identified for larger SkX-2 reciprocal vectors. We therefore refer to this specific realization of the SkX-2 on the lattice as the *lattice* SkX-2. This is illustrated in Fig. 2(c). In contrast, the SkX-2, which forms closer to the ferromagnetic order, in the region of Fig. 1 where the numerical results do not match any of the analytical ground-state energy curves, has a larger unit cell that increases with $\theta$ in a way that is not necessarily commensurate with the underlying triangular lattice. In this regime, we refer to the skyrmion crystal as the *free* SkX-2. A typical representation is given in Fig. 2(d).

## III. ORIGIN OF THE SKYRMION CRYSTAL

At $\theta = \pi$ ($J = -1$ and $J_\chi = 0$), the ground state is ferromagnetic. As $\theta$ decreases, we have numerically found a transition to the SkX-2 at $\theta = 2.4190$. To understand the emergence of this skyrmion order, we consider the continuum expansion of the ferromagnet of the Hamiltonian in Eq. (1), at leading order in the derivatives of the magnetization field. Using integration by parts, the leading terms of the energy density are second order in the derivatives (see Appendix A):

$$h^{(2)} = -\frac{\sqrt{3}}{2}J(\nabla \boldsymbol{m})^2 - 2J_\chi \boldsymbol{m}\cdot(\partial_x \boldsymbol{m} \times \partial_y \boldsymbol{m}). \tag{3}$$

In this continuum limit, the scalar chirality is proportional to the topological charge:

$$Q = \frac{1}{4\pi}\int \boldsymbol{m}\cdot(\partial_x \boldsymbol{m} \times \partial_y \boldsymbol{m})dxdy. \tag{4}$$

This topological quantum number specifies the element in the second homotopy group $\Pi_2(S^2)$ that classifies the different possible wrappings of the magnetization field on a sphere. The latter can be seen as the two-dimensional plane of magnetic configurations through a stereographic projection.

It is known that the leading-order continuum expansion of the exchange interaction admits a set of metastable solutions, including some with nontrivial topological charges $Q \neq 0$; these fulfill the equation [40]:

$$\int (\nabla \boldsymbol{m})^2 dxdy = 8\pi|Q|. \tag{5}$$

Therefore the energy difference between these solutions and the ferromagnetic solution at leading order is

$$E^{(2)}(Q) = -4\pi(\sqrt{3}J|Q| + 2J_\chi Q). \tag{6}$$

In very good agreement with the numerical result, we can conclude that the scalar chirality favors nontrivial configurations with positive topological charges $Q > 0$ and one of these is selected as the ground state, at the expense of the ferromagnetic state ($Q = 0$), when $J_\chi/J = \tan\theta < -\sqrt{3}/2$ or $\theta < \arctan(-\sqrt{3}/2) = 2.4279$. The difference with the numerical value of 2.4190 can be justified due to the large size of the SkX-2 unit cell close to the transition, which is not tractable in the finite-size numerical optimization. In Fig. 3, we plot the spatially averaged charge density $q$, evaluated numerically according to this continuum approximation, i.e.,

$$q = \frac{Q}{A} \approx -\frac{1}{8\pi A}\sum_{(i,j,k)\in\Delta} \boldsymbol{S}_i \cdot (\boldsymbol{S}_j \times \boldsymbol{S}_k), \tag{7}$$

where $A$ is the area of the sample and $a$ is the lattice constant. We observe a continuously increasing charge density from the ferromagnetic transition, consistent with the large SkX-2 unit cell close to the ferromagnet, given the relationship between the SkX-2 reciprocal vectors and the charge density ($q = (|\boldsymbol{X}_i|/\pi)^2/\sqrt{3}$). This smooth evolution is not captured by the leading-order expansion [Eqs. (3)–(6)], which instead yields a sudden jump to an infinitely large charge density at the transition.

## IV. EVOLUTION OF THE TOPOLOGICAL CHARGE

To predict the charge density, we introduce the next-order terms in the derivatives of the magnetization field. These are fourth-order terms in the derivatives because the third-order terms are not invariant under the point-group symmetries of the triangular lattice, e.g., the $\pi$-rotation, which transforms $x \to -x$ and $y \to -y$. Overall, we obtain (see Appendix A):

$$h^{(4)} = h^{(2)} + \frac{\sqrt{3}Ja^2}{32}\left(\partial_x^2 \boldsymbol{m} + \partial_y^2 \boldsymbol{m}\right)^2 + \frac{J_\chi a^2}{4}\boldsymbol{m} \cdot \left(\partial_y \boldsymbol{m} \times \partial_x^3 \boldsymbol{m} + \partial_y^3 \boldsymbol{m} \times \partial_x \boldsymbol{m}\right) + \frac{J_\chi a^2}{4}\boldsymbol{m}$$

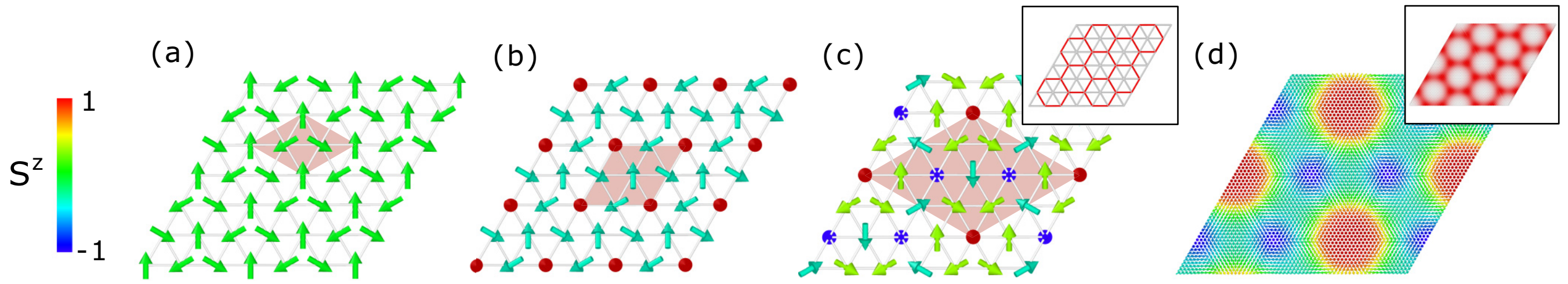

FIG. 2. Schematic view of the spins for (a) the 120°-order, (b) the tetrahedral order, (c) the lattice SkX-2 and (d) the free SkX-2 ($q \approx 0.002/a^2$). The magnetic unit cells are highlighted by the pink quadrilaterals. The inset in panel (c) describes the nearest-neighbor bonds $\boldsymbol{S}_i \cdot \boldsymbol{S}_j$ in the lattice SkX-2 – gray (red) is (anti)ferromagnetic. This demonstrates the lattice symmetry breaking of the lattice SkX-2. The inset in panel (d) shows the distribution of charge density according to Eq. (7), where red and white correspond to positive and zero charge density, respectively. The charge is not localized around different centers like in SkX-1. Nevertheless, every white center lacking charge comes along with half a charge, and there are four of them in one magnetic unit cell.

$$\cdot\,(\partial_{xxy}\boldsymbol{m} \times \partial_x\boldsymbol{m} + \partial_y\boldsymbol{m} \times \partial_{xyy}\boldsymbol{m}) + \frac{J_\chi a^2}{4}\boldsymbol{m} \cdot \big(\partial_{xy}\boldsymbol{m} \times \partial_x^2\boldsymbol{m} + \partial_y^2\boldsymbol{m} \times \partial_{xy}\boldsymbol{m}\big). \tag{8}$$

These higher-order terms introduce the effect of the lattice in the continuum theory, and the triangular lattice constant $a$ appears in the expression. The corresponding energy of the skyrmion solutions can no longer be directly written in terms of the topological charge $Q$, which is dimensionless.

To proceed with evaluating the charge density, we consider the simplest skyrmion unit: a $Q = \pm 1$ skyrmion whose magnetization profile is a metastable solution of the leading-order continuum expansion of the exchange energy and can be expressed, up to a global rotation, as follows [40–42]:

$$\boldsymbol{m}(\boldsymbol{r}) = \frac{1}{(r^2+\lambda^2)}\begin{bmatrix} 2\lambda x \\ -2Q\lambda y \\ (r^2-\lambda^2)\end{bmatrix}, \tag{9}$$

with $\lambda$ being the typical radius of the skyrmion. In the familiar case of a magnetic field along $+z$, the metastable solution can only be rotated about the $z$ axis and the sign of $Q$ is arbitrary, resulting in a distinction between skyrmions ($Q = 1$) and antiskyrmions ($Q = -1$). In contrast, the presence of the SO(3)-symmetric scalar chirality selects the charge but leaves the orientation free.

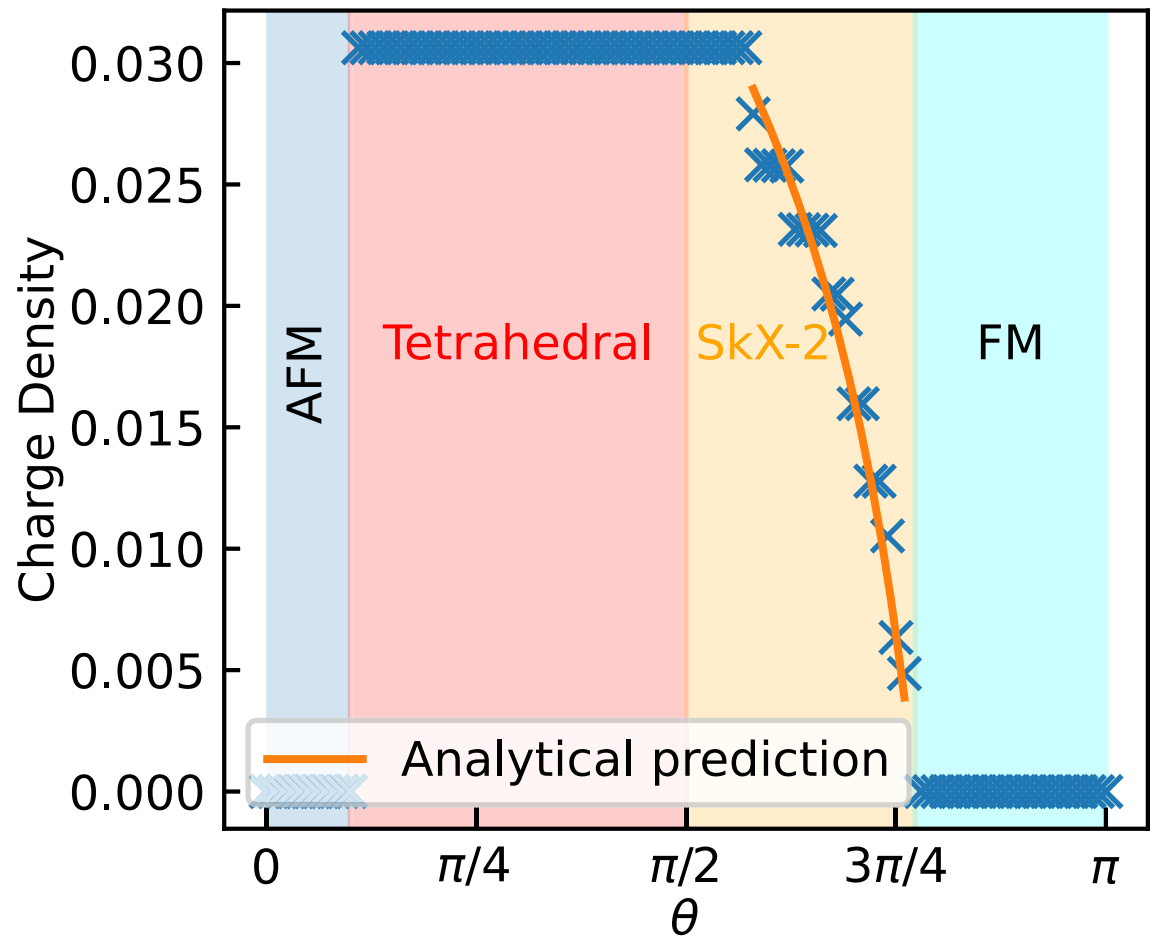

FIG. 3. Topological charge density $q$ as a function of $\theta$ for the ground state, whose energy is given in Fig. 2. The orange line corresponds to the charge density that minimizes the energy in the analytical model, as given by Eq. (12).

By solving the spatial integrals numerically, we evaluate the energy of the $Q = 1$ skyrmion according to Eq. (8):

$$E^{(4)}(Q=1) = E^{(2)}(Q=1) + 2\pi\left(\frac{a}{\lambda}\right)^2\left(\frac{1}{\sqrt{3}}J + 2J_\chi\right). \tag{10}$$

In agreement with the dynamical collapse of skyrmions found in Ref. [41], we see that the leading-order lattice contribution to the ferromagnetic exchange ($J < 0$) favors small skyrmions by becoming more negative with $(a/\lambda)^2$. On the other hand, the lattice contribution to the scalar chirality ($J_\chi > 0$) tends to expand the skyrmion size. In the regime of the SkX-2 ($J_\chi/J < -\sqrt{3}/2$), the lattice contribution to the scalar chirality compensates the one from the exchange, and overall supports the largest skyrmions.

We continue by coarse-graining the problem and assume that the skyrmion crystal is composed of $Q = 1$ skyrmions with a packing fraction $p$. We therefore consider $pA/\pi\lambda^2$ skyrmions. Neglecting the interactions between skyrmions and thus using Eq. (10) as the total energy per skyrmion, we obtain an equation for the energy density as a function of the charge density $q \geqslant 0$ in the form of a Ginzburg-Landau model:

$$e^{(4)}(q) = -4\pi(\sqrt{3}J + 2J_\chi)q + \frac{2\pi^2 a^2}{p}\left(\frac{1}{\sqrt{3}}J + 2J_\chi\right)q^2. \tag{11}$$

Therefore the positive charge density that minimizes the energy in the SkX-2 is given by

$$q = \frac{p}{\pi a^2}\left(\frac{3J + 2\sqrt{3}J_\chi}{J + 2\sqrt{3}J_\chi}\right) = \frac{p}{\pi a^2}\left(\frac{3 + 2\sqrt{3}\tan\theta}{1 + 2\sqrt{3}\tan\theta}\right). \tag{12}$$

This finally predicts theoretically a continuous increase of the charge density from zero at the transition point, in agreement with the numerical results. The exponent $\beta$ describing the growth of the charge density in the SkX-2 order as a function of the distance to the transition $-\sqrt{3}/2 - J_\chi/J$ is 1. In Fig. 3, we plot this prediction with the best fitting $p$ ($p = 0.1106$) for the numerical results of the free SkX-2. The best match between the theory and the numerical results is expected close to the transition where the lattice effect can indeed be treated perturbatively, and the charge density should evolve continuously; however, this is a limit that is not tractable computationally as explained previously.

Moving further away towards the lattice SkX-2, the continuum approach becomes gradually less accurate, as higher-order lattice effects enter in the calculation. The numerical results in Fig. 3 hint at the presence of intermediate plateaus in the charge density, each of which indicates the extended stability (over a finite $\theta$ range) of a configuration with a fixed commensurate SkX-2 reciprocal vector.

Eventually, we observe a saturation in the charge density at $q \approx 0.052/a^2$. This is the lattice SkX-2 [Fig. 2(c)], for which the reciprocal vectors are commensurate and represent half of the $K$ vectors of the Brillouin zone, such that both the size of the SkX-2 unit cell and the scalar chirality energy density are as small as possible. In the lattice SkX-2, the charge density, or the scalar chirality energy density, is homogeneously distributed over all the elementary triangles.

Upon further decrease of $\theta$ below $\theta = \pi/2$, the exchange interaction becomes antiferromagnetic, and the ground state is the tetrahedral order. The scalar chirality energy density remains the same as in the lattice SkX-2. In the two cases, the spins of the triangles point along the directions of three different vertices of a tetrahedron. However, in the tetrahedral order, the nearest-neighbor bonds are all antiferromagnetic ($\boldsymbol{S}_i \cdot \boldsymbol{S}_j < 0$), while the lattice SkX-2 has two ferromagnetic bonds ($\boldsymbol{S}_i \cdot \boldsymbol{S}_j > 0$) for one antiferromagnetic bond, as seen in the inset of Fig. 2(c). This unambiguously justifies the switch between these two orders at $\theta = \pi/2$ when $J$ changes sign.

## V. THERMAL PROPERTIES

To obtain the thermal properties of the system, we use heat-bath Monte Carlo simulations [43]. The update consists of computing the probability density function for a randomly chosen single spin while keeping all other sites fixed. From this distribution, we sample the new spin orientation which is then always accepted. This automatically reduces the variation in spin movement for lower temperatures and allows us to probe a wide range of temperatures. To improve convergence behavior and strengthen decorrelation, we used replica exchange steps [44–47] and over-relaxation [47].

### A. Lattice SkX-2

We are primarily interested in the thermal properties of the lattice SkX-2, for which we can observe a clear peak in the heat capacity that diverges with the system size. To study the nature of this transition, we define an order parameter for the lattice SkX-2, which uses the alternating sign of the bond energy around every second spin, as seen in the inset of Fig. 2(c). We define

$$P(i) = \boldsymbol{S}_i \cdot (\boldsymbol{S}_j + \boldsymbol{S}_l + \boldsymbol{S}_n - \boldsymbol{S}_k - \boldsymbol{S}_m - \boldsymbol{S}_o), \tag{13}$$

where $j$, $k$, $l$, $m$, $n$, and $o$ are the ordered indices of the nearest neighbors of lattice site $i$. The order parameter is simply given by the magnitude of $P(i)$ in momentum space at $\boldsymbol{K}_1 = \frac{2\pi}{3a}(2, 0)$ and $\boldsymbol{K}_2 = \frac{2\pi}{3a}(1, \sqrt{3})$, i.e.,

$$M_{\mathrm{SkX-2}} = |\tilde{P}(\boldsymbol{K}_1)| + |\tilde{P}(\boldsymbol{K}_2)|. \tag{14}$$

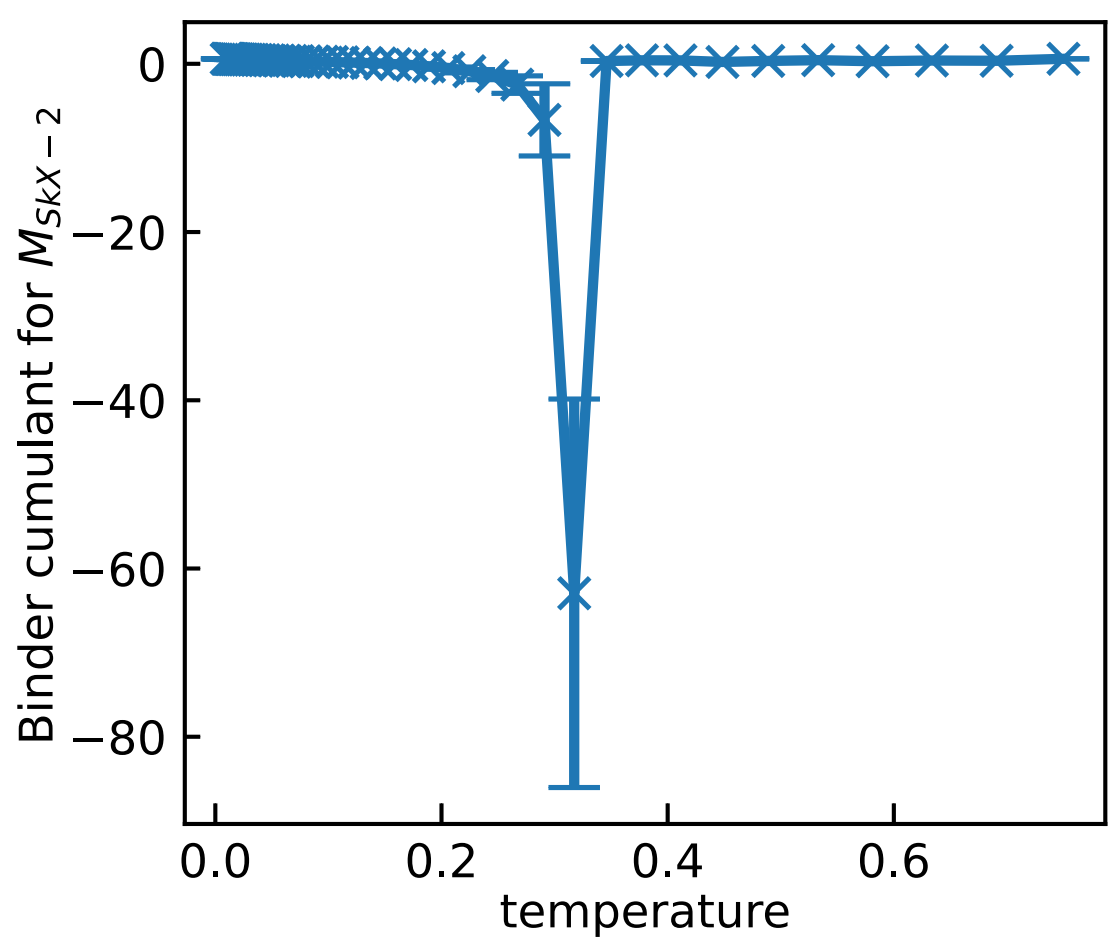

FIG. 4. Binder parameter for the lattice SkX-2 order parameter on the 240 × 240-spin triangular lattice for $\theta = 51\pi/100$.

We can then compute the Binder cumulant of this order parameter as

$$U = 1 - \frac{\langle M_{\mathrm{SkX-2}}^4 \rangle}{3\langle M_{\mathrm{SkX-2}}^2 \rangle^2}. \tag{15}$$

We observe a sharp dip of the Binder parameter as a function of the temperature in Fig. 4, revealing the coexistence of an ordered and disordered phase at the critical temperature, underlying a first-order phase transition. This transition corresponds to the spin lattice translation symmetry breaking related to the bond energy distribution [inset of Fig. 2(c)].

### B. Free SkX-2

The situation in the regime of the free SkX-2 is more subtle. Investigation of the heat capacity at an arbitrary point close to the ferromagnetic transition in Fig. 5(a) indicates the presence of a bump followed by a sharp peak as temperature decreases. The peak does not seem to diverge with system size and should therefore ultimately be related to a distinct continuous feature.

In line with previous studies on the thermal properties of SkX-1 [48–54], we investigate the possible existence of KTHNY melting on the free SkX-2. In the absence of a substrate, KTHNY theory describes the melting processes of particles in the plane and is distinguished by the occurrence of two subsequent continuous transitions. The first at higher temperature results in a quasi-long-range orientational order (a hexatic phase), characterized by the polynomial decay of the orientational order parameter's correlation function. The second produces a quasi-long-range translational order (a floating solid phase), characterized by the polynomial decay of the translational order parameter's correlation function. In the presence of a substrate, played in our case by the spin triangular lattice on the skyrmion texture, the situation is altered [34]: for a strong potential induced by the substrate, the two original continuous transitions are replaced by a unique finite-order phase transition, as observed for the lattice SkX-2. For the free SkX-2, however, it is known that the presence of any weak substrate with the same discrete rotational symmetry as the crystal makes the quasi-long-range orientational order irrelevant, or, in other words, emerging continuously from infinite temperature. As the triangular spin lattice shares the same sixfold orientation symmetry as the SkX-2, we therefore cannot expect a sudden formation of quasi-long-range orientational order.

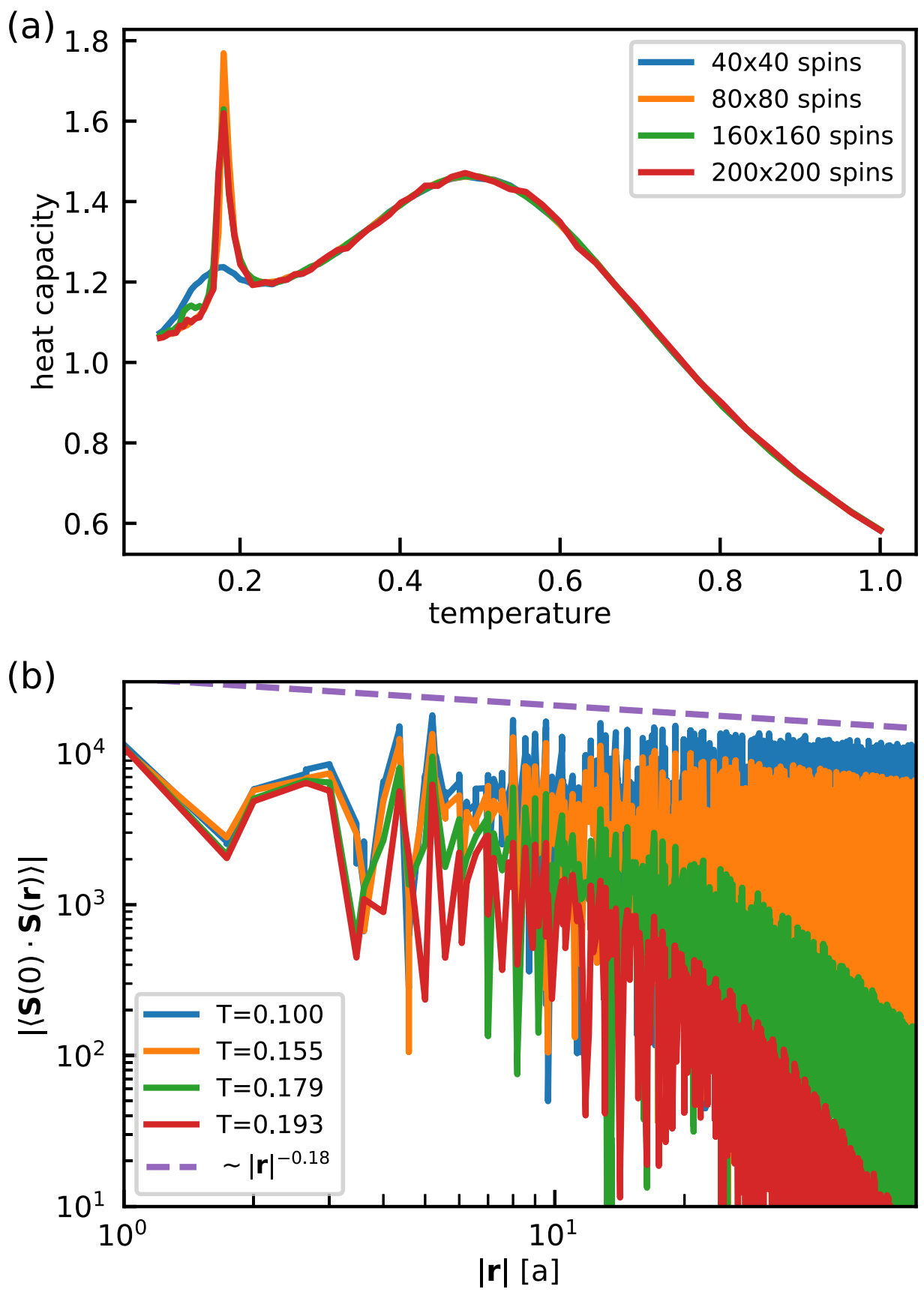

FIG. 5. (a) Heat capacity as a function of temperature for different system sizes at $\theta = 1.96$. The sharp peak that does not diverge with system size at $T \approx 0.17$ indicates the presence of a distinct continuous feature. (b) Spin-spin correlation function at $\theta = 1.96$ for different temperatures. The data up to a separation distance $|\boldsymbol{r}| = 70a$ indicate a change of behavior from exponential to power law at $T \approx 0.17$, suggesting the appearance of quasi-long-range order persisting at least down to $T = 0.10$.

With this in mind, we examine the evolution of the spin-spin correlation function with temperature in Fig. 5(b) and observe the appearance of quasi-long-range order coinciding with the sharp peak in the heat capacity. This supports the existence of a continuous transition to a floating solid, given that the spin correlation function relays the same information as the correlation of a coarse-grained translational order parameter constructed ad hoc from the charge density—with the latter containing the information about the position of the effective particles. The exponent $\eta$ for the correlation decay $\sim r^{-\eta}$ is approximately 0.18, much smaller than the upper bound of $1/3$ considered for a floating solid of skyrmions in Ref. [51].

To test the presence of a floating solid further, we recall that its melting in KTHNY theory is described by the separation of dislocation pairs, for which the dislocations are bound five- and sevenfold defect environments within the defect-free sixfold vertices of the hexagonal particle arrangement. In the free SkX-2, we can also recognize an emergent hexagonal structure from the regions without charge density, as seen from the white areas in the inset of Fig. 2(d). Based on this observation, we set out to reconstruct and analyze the emergent hexagonal structure from the thermal spin configurations. For that, we relax the thermal spin configuration with gradient descent to a fixed norm of the residual effective field and identify the vertices of the hexagonal lattice by the centroids of the clusters made of the spin-lattice triangular plaquettes with a scalar chirality below the median. The resulting emergent hexagonal lattices, illustrated in Fig. 6, clearly demonstrate the presence of many distant dislocations just above the transition temperature, which almost completely disappear below the transition temperature. To track their evolution more broadly across the transition temperature, we plot the density of five- and sevenfold coordinated environments, in addition to four- and eightfold environments for completeness, in Fig. 6(c).

This analysis confirms the relevance of the dislocations in the continuous thermal transition in the free SkX-2, and therefore the presence of KTHNY physics, taking place here as a single topological transition to a floating solid of skyrmions. The almost complete absence of dislocations below the transition temperature, despite the theory predicting bound pairs, is interpreted as the effect of the relaxation: some defect structures, potentially including bound dislocation pairs, are not necessarily immune to the gradient descent. This is usually a feature of structures that are not protected by an energy barrier larger than the thermal energy scale.

### C. Tetrahedral order

To complete our analysis of the thermal properties of the model, we discuss and investigate the possible thermal transitions in the other phases. According to the Mermin-Wagner theorem [31,32] and the absence of further ground-state order with translational symmetry breaking, no spontaneous symmetry-breaking transition is expected beyond that of the lattice SkX-2.

Nevertheless, despite the three-dimensional nature of the spins in this model, it is known that a type of Berezinskii-Kosterlitz-Thouless (BKT) transition can still occur on the 120°-order [35,55]. This is attributed to the configuration space of the order, whose order parameter is constituted by three spins of the elementary triangle, i.e., three independent spins of the unit cell depicted in Fig. 2(a). Due to the Heisenberg nature of the spins, the arrangement of these three spins in the 120°-order is isomorphic to the projective space $P_3$, or equivalently, the parameter space of the SO(3) group in the three-dimensional representation. However, the first homotopy group of $P_3$ is equal to $\mathbb{Z}_2$, i.e., $\pi_1(P_3) = \mathbb{Z}_2$, indicating the presence of nontrivial point defects. In the magnetization, this corresponds to the existence of a trivial and a nontrivial winding configuration of the order parameter in the plane, with the nontrivial arrangement corresponding to a vortex. As vortices only appear in pairs at low temperature, there is a temperature at which they dissociate, leading to a topological phase transition. However, in contrast to the BKT transition of the XY model, there is no signature of this topological phase transition in the decay of the spin-correlation functions.

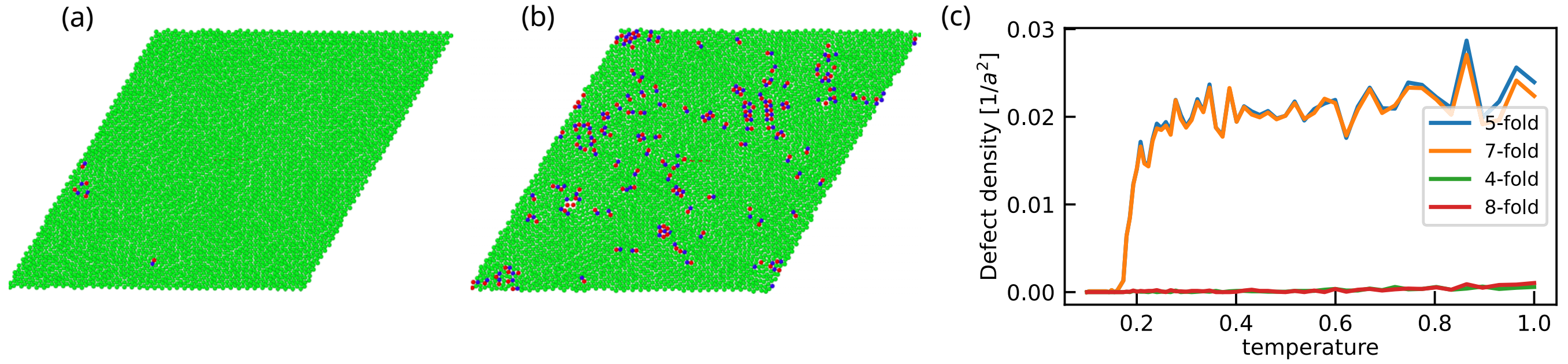

FIG. 6. Emergent hexagonal structure made of charge-free centers at $\theta = 1.96$ obtained from spin configurations relaxed with gradient descent method at (a) $T = 0.155$ and (b) 0.179. Green vertices are sixfold coordinated, while blue and red are defective five- and sevenfold coordinated vertices, respectively. Coordination numbers are obtained from a Voronoi tessellation. (c) Density of vertices in the emergent hexagonal structure. At $T \approx 0.17$, there is an abrupt joint jump of five- and sevenfold coordinated defects, which by pairs form dislocations.

In the present analysis, we do not intend to provide further evidence for a $\mathbb{Z}_2$-vortex topological transition in the 120°-order beyond what is presented in Refs. [35,55]. Nevertheless, by acknowledging the conditions supporting this transition, we also recognize its potential relevance to the tetrahedral order. This is because the four-spin order parameter of the tetrahedral order [see Fig. 9(b)] also transforms like the vector representation of SO(3). With this in mind, we start by investigating the low-temperature spin configurations and unambiguously find the presence of defect pairs. In Figs. 7(a)–7(b), we show an example in which the defect pair is most clearly observed by inspecting a single sublattice.

To determine whether these defects are vortex pairs, we examine the scaling of their energy with respect to the Hamiltonian parameters and compare it with the theoretical expectation. On the one hand, we can use mART [56], a magnetic energy landscape exploration method, to access different metastable configurations where the two defects are separated by varying distances. We plot the resulting energy as a function of pair separation in Fig. 7(c). On the other hand, we can analytically evaluate the energy of a vortex defined by a scalar field $\phi(\boldsymbol{r})$ superposed to the spin orientation in an arbitrary plane of the spin space. The field $\phi$ is directly given by the polar angle $\alpha$, measured from the vortex center: $\phi(r, \alpha) = \alpha$. By doing so (see Appendix B), the energy stored between distances $l$ and $L$ from the center of the vortex is given by

$$E_{\text{vortex}} = \frac{\sqrt{3}\pi}{12} \ln(L/l) F(J, J_\chi), \tag{16}$$

where for the tetrahedral order,

$$F(J, J_\chi) \approx 2.7J + 18.5J_\chi. \tag{17}$$

We can now verify whether the defect pair energy density, evaluated numerically as the difference between the energy density of the configuration $e$ and the tetrahedral-order energy density $e_0$, scales consistently with the vortex energy by rescaling the energy by $F(J, J_\chi)$ in Fig. 7(d). Since the resulting curves overlap, we can confirm that the low-temperature phase above the tetrahedral ground state is composed of vortex pairs, providing strong evidence for a $\mathbb{Z}_2$-vortex topological transition, as in the 120°-order.

## VI. DISCUSSION AND CONCLUDING REMARKS

The Heisenberg model augmented by a scalar chirality term on the triangular lattice is an interesting model for hosting complex magnetic textures in the ground state, in particular the skyrmion crystal SkX-2, and supporting both a conventional first-order phase transition and less conventional transitions related to KTHNY theory and the unbinding of topological vortices. Our combined numerical and analytical analysis has shed light on these various phenomena.

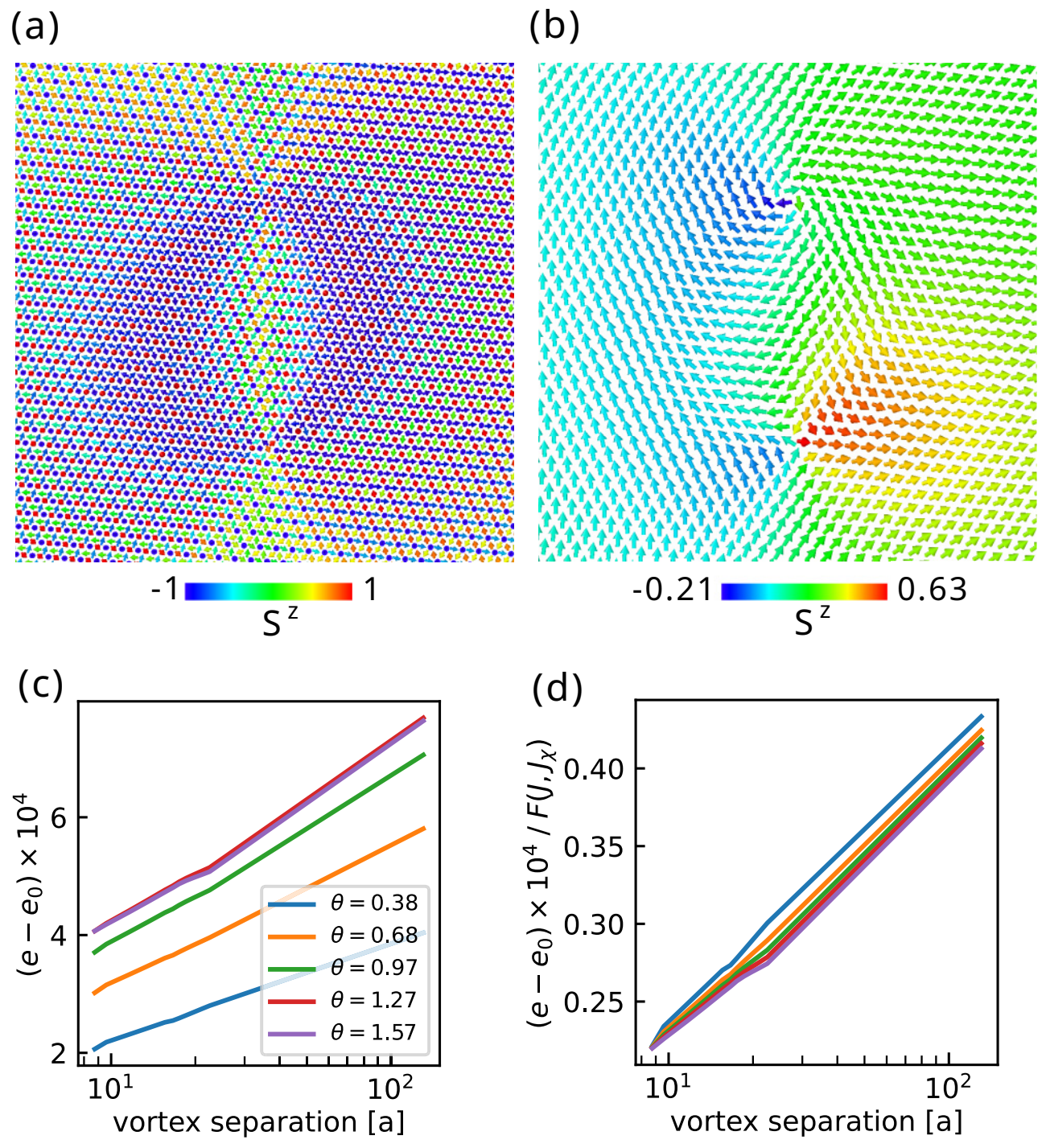

FIG. 7. (a) Defect pair configuration found at $T = 0.01$ and $\theta = 0.38$. (b) Same configuration plotted only with spins belonging to one sublattice (one spin per tetrahedral-order unit cell). (c) Defect pair energy density $e - e_0$ as a function of separation evaluated from different metastable configurations. (d) Defect pair energy renormalized by analytical vortex energy factor $F(J, J_\chi)$ defined in Eq. (17).

The presence of a topologically nontrivial solution in the ground state can be seen as the consequence of the scalar chirality term, which, in the continuum limit, is directly proportional to the topological charge, while the exchange interaction already admits this solution as a metastable state. The absence of the magnetic field in this SO(3)-symmetric model allows for the emergence of a chiral SkX-2 ground state presenting a continuous symmetry in spin space and bearing no magnetization at the Γ-point in reciprocal space, in contrast to the usual SkX-1 arrangement of (anti)skyrmions.

We found that the SkX-2 allows for a monotonic, gradual increase of the topological charge density between the ferromagnetic and tetrahedral ground states, notably through the free SkX-2 regime where the large unit cell does not necessarily have to be commensurate with the underlying spin lattice. Nevertheless, our numerical analysis also hints at the existence of some commensurate ground states when approaching the lattice limit of the SkX-2, as suggested by plateaus in the charge density in Fig. 3.

The continuous emergence of the charge density in the free SkX-2, starting at zero in the trivial ferromagnetic order, can be understood due to the lattice contribution, which favors the smoothest magnetic texture at the transition. Assuming noninteracting localized $Q = 1$ skyrmions and a single fitting parameter corresponding to the packing fraction, we found good quantitative agreement with the evolution of the charge density evaluated numerically. The fitted packing fraction is 0.1106 for skyrmions with radius defined as $\lambda$ in Eq. (9). As a comparison, the highest-density lattice packing of circles (hexagonal lattice) is $\pi/2\sqrt{3} \approx 0.9069$. This large difference with the optimal packing fraction can a priori be understood from the fact that skyrmions are usually interacting in a complex way, making it difficult to establish their true effective radius [57,58]. More importantly, this result should be rationalized from the fact that the SkX-2 is not a texture made of well-localized skyrmions, but rather of a topological charge spread along the edges and vertices of a honeycomb lattice [Fig. 2(d)]. Even if the charge were localized at the vertices of the emergent honeycomb lattice, it would only be 1/4 per node, still breaking down the representations of individual integer $Q = 1$ skyrmions, despite its validity in predicting the continuous transition from the ferromagnet.

Our analysis of the thermal behavior using Monte Carlo simulations has shown evidence for a very different type of thermal transition between the lattice and free SkX-2. While the lattice SkX-2 exhibits only a discrete lattice translational symmetry, resulting in a first-order phase transition as found with the Binder cumulant method, the free SkX-2 has a continuous translation symmetry and a behavior in perfect agreement with KTHNY theory for two-dimensional particle systems. In particular, the size-independent peak in the heat capacity was associated with the emergence of a KTHNY floating solid. Although KTHNY physics has already been considered in various SkX-1 systems [48–54], the most relevant evidence often relies on constructing an arbitrary skyrmion-to-point-particle mapping, from which the spatial correlation of the key translational (or orientational) order parameter is determined. In our methodology, we directly detected the transition to quasi-long-range order from the spin-spin correlation function. Conversely, we still used the skyrmion-to-particle mapping on relaxed SkX-2 thermal configurations to identify the charge-free centers and confirm the relevant dislocation physics.

The dislocations identified are highly mobile defects. At $\theta = 1.96$, a separate analysis with mART [56] has established that the typical energy barrier for the relative motion of a dislocation pair is about three orders of magnitude smaller than the thermal energy scale at the transition ($T_c \approx 0.17$). This implies that dislocations diffuse well below the transition, as already observed experimentally in the melting of a different skyrmion lattice in Ref. [59]. However, as $\theta$ is reduced, the effect of the underlying pinning potential created by the spin lattice becomes stronger, which should gradually impede the dislocation mobility.

In experiments, characterizing the SkX-2 state should *a priori* be possible via direct spatial imaging or indirect transport probes. In particular, a prominent signature is the expected topological Hall conductivity resulting from current deflection, whose direction is dictated by the sign of $J_\chi$ in our model and has been explicitly visualized as itinerant electron current loops in Ref. [14]. Nevertheless, one must keep in mind that such experimental observations will be highly sensitive to the level of structural disorder. More specifically, one needs to understand how quenched disorder affects the stability of the SkX-2 thermal phases, for instance by adding a random pinning potential to the magnetic texture [60]. Quenched disorder fundamentally alters the spatial symmetries of the system: it breaks both the global sixfold point-group symmetry and the discrete translational symmetry of the lattice Hamiltonian. The sixfold symmetry originally supports the continuous growth, from infinite temperature, of a sixfold orientational order. Consequently, this orientational order should be highly robust against weak quenched disorder, as it is against thermal fluctuations. In contrast, the discrete translational symmetry of the lattice yields a floating solid at low temperatures characterized by power-law positional correlations.

Because the floating solid is fragile due to its quasi-long-range nature, even weak disorder is expected to act as a relevant perturbation that will lower, but not directly cancel, the floating-solid transition, as seen in weakly disordered colloidal monolayers, vortex lattices and skyrmion lattices [59,61,62]. The floating solid could therefore be partially replaced by the (SkX-2) paramagnet, which exhibits a sixfold orientational symmetry and a dense fraction of unbounded dislocation pairs. In the remaining floating solid phase, the shape of the magnetic texture will still be locally distorted due to the random pinning [63].

To conclude, the studied model can also be seen as a natural extension to probe the 120°-order, whose evolution under thermal fluctuations is subtle [55]. In particular, investigating the low-temperature configurations above the tetrahedral state, which replaces the 120°-order for a large scalar chirality, we unambiguously identified the presence of $\mathbb{Z}_2$ vortex pairs. Their presence seems to be at the origin of the same $\mathbb{Z}_2$ vortex topological transition as above the 120°-state. With this in mind, one could start exploring this intriguing thermal

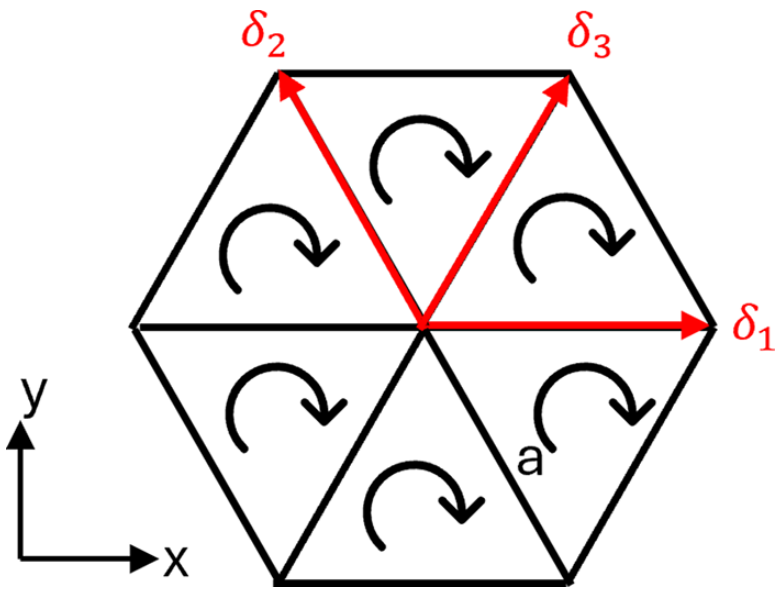

FIG. 8. Definition of the three unit lattice vectors $\delta_1$, $\delta_2$, and $\delta_3$ and illustration of the clockwise permutation for the scalar chirality on every elementary triangle.

transition on new platforms supporting a tetrahedral order, such as Mn monolayers [64,65].

## ACKNOWLEDGMENT

The present work was supported by the Swiss National Science Foundation under Grant No. 200021-196970.

## DATA AVAILABILITY

There are no publicly available research data or software supporting this manuscript. Requests for further information or data should be sent to the authors.

## APPENDIX A: CONTINUUM EXPANSIONS ON THE FERROMAGNET

We define the three lattice vectors according to Fig. 8. We start by expanding the exchange Hamiltonian, which can be rewritten as

$$\mathcal{H}_J = \frac{1}{2}\sum_i \boldsymbol{S}_i \cdot \big(\boldsymbol{S}_{i+\delta_1} + \boldsymbol{S}_{i-\delta_1} + \boldsymbol{S}_{i+\delta_2} + \boldsymbol{S}_{i-\delta_2} + \boldsymbol{S}_{i+\delta_3} + \boldsymbol{S}_{i-\delta_3}\big). \tag{A1}$$

By additionally defining the following derivatives:

$$\nabla_1 = a\partial_x,\ \nabla_2 = \frac{a}{2}(-\partial_x + \sqrt{3}\partial_y) \text{ and } \nabla_3 = \frac{a}{2}(\partial_x + \sqrt{3}\partial_y), \tag{A2}$$

we can expand the ferromagnet including terms up to fourth order in the derivatives. Up to a constant, it yields

$$\mathcal{H}_J^{(4)} = \frac{1}{2}\sum_i \boldsymbol{S}_i \cdot \big(\nabla_1^2 + \nabla_2^2 + \nabla_3^2\big)\boldsymbol{S}_i + \frac{1}{24}\sum_i \boldsymbol{S}_i \cdot \big(\nabla_1^4 + \nabla_2^4 + \nabla_3^4\big)\boldsymbol{S}_i. \tag{A3}$$

By explicitly inserting the derivatives, we obtain

$$\mathcal{H}_J^{(4)} = \frac{3a^2}{4}\sum_i \boldsymbol{S}_i \cdot \big(\partial_x^2 + \partial_y^2\big)\boldsymbol{S}_i + \frac{3a^4}{64}\sum_i \boldsymbol{S}_i \cdot \big(\partial_x^2 + \partial_y^2\big)^2\boldsymbol{S}_i. \tag{A4}$$

We make the continuum transformation: $\sum_i f(r_i) = \sum_i 2/(\sqrt{3}a^2)\int_{\text{cell } i} f(r_i) \quad dxdy \to 2/(\sqrt{3}a^2)\int f(r) \quad dxdy$ to write the exchange energy density:

$$h_J^{(4)} = \frac{\sqrt{3}}{2}\boldsymbol{m}\cdot\big(\partial_x^2 + \partial_y^2\big)\boldsymbol{m} + \frac{\sqrt{3}a^2}{32}\boldsymbol{m}\cdot\big(\partial_x^2 + \partial_y^2\big)^2\boldsymbol{m}. \tag{A5}$$

Using integration by parts, we finally get

$$h_J^{(4)} = -\frac{\sqrt{3}}{2}[(\partial_x\boldsymbol{m})^2 + (\partial_y\boldsymbol{m})^2] + \frac{\sqrt{3}a^2}{32}\big(\partial_x^2\boldsymbol{m} + \partial_y^2\boldsymbol{m}\big)^2. \tag{A6}$$

For the scalar chirality energy, we account directly for the lattice symmetry in the expansion by writing

$$\mathcal{H}_{J_\chi} = \frac{1}{3}\sum_i \sum_{(i,j,k)\in\Delta} \boldsymbol{S}_i \cdot (\boldsymbol{S}_j \times \boldsymbol{S}_k), \tag{A7}$$

where the second sum is taken over six triangles of the vertex $i$ as depicted in Fig. 8. We can rewrite the Hamiltonian:

$$\mathcal{H}_{J_\chi} = \frac{1}{3}\sum_i \boldsymbol{S}_i \cdot \big(\boldsymbol{S}_{i+\delta_2} \times \boldsymbol{S}_{i+\delta_3} + \boldsymbol{S}_{i-\delta_2} \times \boldsymbol{S}_{i-\delta_3} + \boldsymbol{S}_{i+\delta_3} \times \boldsymbol{S}_{i+\delta_1} + \boldsymbol{S}_{i-\delta_3} \times \boldsymbol{S}_{i-\delta_1} + \boldsymbol{S}_{i+\delta_1} \times \boldsymbol{S}_{i-\delta_2} + \boldsymbol{S}_{i-\delta_1} \times \boldsymbol{S}_{i+\delta_2}\big). \tag{A8}$$

As with the exchange Hamiltonian, we can start by expanding the ferromagnet up to second order in the derivatives:

$$\begin{aligned}\mathcal{H}_{J_\chi}^{(2)} &= \frac{2}{3}\sum_i \boldsymbol{S}_i \cdot (\nabla_2\boldsymbol{S}_i \times \nabla_3\boldsymbol{S}_i + \nabla_3\boldsymbol{S}_i \times \nabla_1\boldsymbol{S}_i - \nabla_1\boldsymbol{S}_i \times \nabla_2\boldsymbol{S}_i)\\ &= -\sqrt{3}a^2\sum_i \boldsymbol{S}_i \cdot (\partial_x\boldsymbol{S}_i \times \partial_y\boldsymbol{S}_i).\end{aligned} \tag{A9}$$

We now apply the continuum transformation:

$$h_{J_\chi}^{(2)} = -2\boldsymbol{m}\cdot(\partial_x\boldsymbol{m} \times \partial_y\boldsymbol{m}). \tag{A10}$$

Up to fourth order in the derivatives, we obtain

$$\begin{aligned}\mathcal{H}_{J_\chi}^{(4)} = \mathcal{H}_{J_\chi}^{(2)} &+ \frac{1}{9}\sum_i \boldsymbol{S}_i \cdot \big[\nabla_2^3\boldsymbol{S}_i \times (\nabla_3 + \nabla_1)\boldsymbol{S}_i\big]\\ &+ \frac{1}{9}\sum_i \boldsymbol{S}_i \cdot \big[\nabla_3^3\boldsymbol{S}_i \times (\nabla_1 - \nabla_2)\boldsymbol{S}_i\big]\\ &- \frac{1}{9}\sum_i \boldsymbol{S}_i \cdot y\big[\nabla_1^3\boldsymbol{S}_i \times (\nabla_2 + \nabla_3)\boldsymbol{S}_i\big]\\ &+ \frac{1}{6}\sum_i \boldsymbol{S}_i \cdot \big[\nabla_2^2\boldsymbol{S}_i \times \nabla_3^2\boldsymbol{S}_i + \nabla_3^2\boldsymbol{S}_i \times \nabla_1^2\boldsymbol{S}_i\\ &+ \nabla_1^2\boldsymbol{S}_i \times \nabla_2^2\boldsymbol{S}_i\big],\end{aligned} \tag{A11}$$

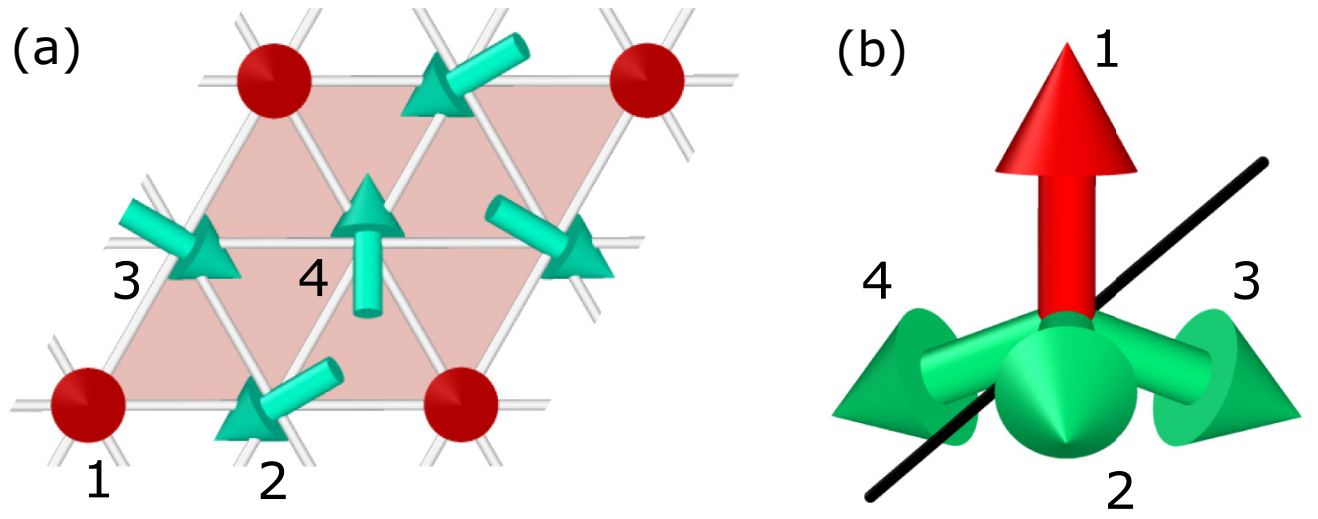

FIG. 9. (a) Unit cell of tetrahedral order [Fig. 2(b)] composed of four different spins. (b) The orientations of these spins form a tetrahedron within which the vortex develops in a specific plane, for instance perpendicular to the black line.

which is equal to

$$\mathcal{H}_{J_\chi}^{(4)} = \mathcal{H}_{J_\chi}^{(2)} + \frac{\sqrt{3}a^4}{8}\sum_i \boldsymbol{S}_i \cdot \left(\partial_y \boldsymbol{S}_i \times \partial_x^3 \boldsymbol{S}_i + \partial_y^3 \boldsymbol{S}_i \times \partial_x \boldsymbol{S}_i\right) + \frac{\sqrt{3}a^4}{8}\sum_i \boldsymbol{S}_i \cdot (\partial_{xxy}\boldsymbol{S}_i \times \partial_x \boldsymbol{S}_i + \partial_y \boldsymbol{S}_i \times \partial_{xyy}\boldsymbol{S}_i) + \frac{\sqrt{3}a^4}{8}\sum_i \boldsymbol{S}_i \cdot \left(\partial_{xy}\boldsymbol{S}_i \times \partial_x^2 \boldsymbol{S}_i + \partial_y^2 \boldsymbol{S}_i \times \partial_{xy}\boldsymbol{S}_i\right). \tag{A12}$$

In the continuum limit, we obtain

$$h_{J_\chi}^{(4)} = h_{J_\chi}^{(2)} + \frac{a^2}{4}\boldsymbol{m} \cdot \left(\partial_y \boldsymbol{m} \times \partial_x^3 \boldsymbol{m} + \partial_y^3 \boldsymbol{m} \times \partial_x \boldsymbol{m}\right) + \frac{a^2}{4}\boldsymbol{m} \cdot (\partial_{xxy}\boldsymbol{m} \times \partial_x \boldsymbol{m} + \partial_y \boldsymbol{m} \times \partial_{xyy}\boldsymbol{m}) + \frac{a^2}{4}\boldsymbol{m} \cdot \left(\partial_{xy}\boldsymbol{m} \times \partial_x^2 \boldsymbol{m} + \partial_y^2 \boldsymbol{m} \times \partial_{xy}\boldsymbol{m}\right). \tag{A13}$$

## APPENDIX B: VORTICES ON THE TETRAHEDRAL ORDER

To evaluate the vortex energy on the tetrahedral order, we need to take into account the enlarged unit cell (u.c.) made of four distinct spins according to Fig. 9(a):

$$\mathcal{H} = 2J\sum_{\text{u.c.}}[\boldsymbol{S}_1 \cdot \boldsymbol{S}_2 + \boldsymbol{S}_3 \cdot \boldsymbol{S}_4 + \boldsymbol{S}_1 \cdot \boldsymbol{S}_3 + \boldsymbol{S}_2 \cdot \boldsymbol{S}_4 + \boldsymbol{S}_2 \cdot \boldsymbol{S}_3 + \boldsymbol{S}_1 \cdot \boldsymbol{S}_4] + 2J_\chi \sum_{\text{u.c.}}[\boldsymbol{S}_1 \cdot (\boldsymbol{S}_3 \times \boldsymbol{S}_2) + \boldsymbol{S}_2 \cdot (\boldsymbol{S}_3 \times \boldsymbol{S}_4) + \boldsymbol{S}_2 \cdot (\boldsymbol{S}_4 \times \boldsymbol{S}_1) + \boldsymbol{S}_1 \cdot (\boldsymbol{S}_4 \times \boldsymbol{S}_3)]. \tag{B1}$$

We assume the existence of a vortex as a perturbation in the angle difference between neighboring spins in an arbitrary plane. We decompose the spins with their components in and perpendicular to this plane ($\boldsymbol{S}_i = \boldsymbol{S}_i^{||} + \boldsymbol{S}_i^{\perp}$) such that the typical scalar chirality term becomes

$$\boldsymbol{S}_1 \cdot (\boldsymbol{S}_2 \times \boldsymbol{S}_3) = \sum_{\varepsilon_{ijk}=1} \boldsymbol{S}_i^{\perp} \cdot (\boldsymbol{S}_j^{||} \times \boldsymbol{S}_k^{||}) = \sum_{\varepsilon_{ijk}=1} |\boldsymbol{S}_i^{\perp}| \sin(\psi_{jk} + \nabla_{jk}\phi) \approx \sum_{\varepsilon_{ijk}=1} |\boldsymbol{S}_i^{\perp}| \left[\sin(\psi_{jk}) + \cos(\psi_{jk})(\boldsymbol{r}_{jk} \cdot \nabla\phi) - \frac{1}{2}\sin(\psi_{jk})(\boldsymbol{r}_{jk} \cdot \nabla\phi)^2\right]. \tag{B2}$$

The angles $\psi_{jk}$ and $\phi(\boldsymbol{r})$ are respectively the unperturbed angle difference in the vortex plane between spin $j$ and $k$, and the scalar field associated with the superposition of a vortex. We can integrate this contribution in the continuum limit:

$$\sum_{\text{u.c.}} \boldsymbol{S}_1 \cdot (\boldsymbol{S}_2 \times \boldsymbol{S}_3) = |\boldsymbol{S}_i^{\perp}| \frac{\sqrt{3}}{6a^2}\int d\boldsymbol{r}^2 \sum_{\varepsilon_{ijk}=1}\left[\sin(\psi_{jk}) + \cos(\psi_{jk})(\boldsymbol{r}_{jk} \cdot \nabla\phi) - \frac{1}{2}\sin(\psi_{jk})(\boldsymbol{r}_{jk} \cdot \nabla\phi)^2\right]. \tag{B3}$$

In polar coordinates $(r, \alpha)$ centered at the vortex, and for a vortex $\phi(\alpha) = \alpha$ corresponding to $\boldsymbol{r}_{jk} \cdot \nabla\phi = \frac{1}{r}\frac{\partial\phi}{\partial\alpha}(\boldsymbol{r}_{jk} \cdot \boldsymbol{e}_\alpha) = \cos(\alpha)/r$, we obtain a scalar energy contribution for $l < r < L$:

$$\sum_{\text{u.c.}} \boldsymbol{S}_1 \cdot (\boldsymbol{S}_2 \times \boldsymbol{S}_3) \approx \frac{\sqrt{3}}{2a^2}\overline{\boldsymbol{S}_1} \cdot (\overline{\boldsymbol{S}_2} \times \overline{\boldsymbol{S}_3})\pi \times \left[L^2 - l^2 - \frac{a^2}{2}\ln(L/l)\right], \tag{B4}$$

where $\overline{\boldsymbol{S}_i}$ are the unperturbed spins of the tetrahedral order.

Similarly, we can expand the typical exchange term under a vortex by writing

$$\boldsymbol{S}_1 \cdot \boldsymbol{S}_2 = \boldsymbol{S}_1^{\perp} \cdot \boldsymbol{S}_2^{\perp} + \boldsymbol{S}_1^{||} \cdot \boldsymbol{S}_2^{||} = \boldsymbol{S}_1^{\perp} \cdot \boldsymbol{S}_2^{\perp} + \cos(\psi_{12} + \boldsymbol{r}_{12} \cdot \nabla\phi) \approx \boldsymbol{S}_1^{\perp} \cdot \boldsymbol{S}_2^{\perp} + \cos(\psi_{12}) - \sin(\psi_{12})(\boldsymbol{r}_{12} \cdot \nabla\phi) - \frac{1}{2}\cos(\psi_{12})(\boldsymbol{r}_{12} \cdot \nabla\phi)^2, \tag{B5}$$

and solving the same integrals as in Eq. (B3) to obtain

$$\sum_{\text{u.c.}} \boldsymbol{S}_1 \cdot \boldsymbol{S}_2 \approx \frac{\sqrt{3}}{6a^2}\pi\left[(\overline{\boldsymbol{S}_1} \cdot \overline{\boldsymbol{S}_2})(L^2 - l^2) - (\overline{\boldsymbol{S}_1^{||}} \cdot \overline{\boldsymbol{S}_2^{||}})\frac{a^2}{2}\ln(L/l)\right]. \tag{B6}$$

In total, we can write the vortex energy by gathering terms like Eqs. (B4) and (B6) into Eq. (B1) and subtracting the ground-state energy:

$$E_{\text{vortex}} = \frac{\sqrt{3}\pi}{12} \ln(L/l) F(J, J_\chi), \tag{B7}$$

where

$$\begin{aligned} F(J, J_\chi) = &-2J[\overline{S_1^{||}} \cdot \overline{S_2^{||}} + \overline{S_3^{||}} \cdot \overline{S_4^{||}} + \overline{S_1^{||}} \cdot \overline{S_3^{||}} + \overline{S_2^{||}} \cdot \overline{S_4^{||}} \\ &+ \overline{S_2^{||}} \cdot \overline{S_3^{||}} + \overline{S_1^{||}} \cdot \overline{S_4^{||}}] - 6J_\chi[\overline{S_1} \cdot (\overline{S_3} \times \overline{S_2}) \\ &+ \overline{S_2} \cdot (\overline{S_3} \times \overline{S_4}) + \overline{S_2} \cdot (\overline{S_4} \times \overline{S_1}) \\ &+ \overline{S_1} \cdot (\overline{S_4} \times \overline{S_3})]. \end{aligned} \tag{B8}$$